\documentclass[conference]{IEEEtran}
\usepackage{cite}

\usepackage{graphicx}

\usepackage{algorithm}
\usepackage{algpseudocode}

\usepackage[table,xcdraw]{xcolor}

\begin{document}
%
\title{Automated Prototype for Asteroids Detection}



%
\author{\IEEEauthorblockN{Denisa Cop\^{a}ndean\IEEEauthorrefmark{1}, 
Ovidiu V\u{a}duvescu\IEEEauthorrefmark{3}, Dorian Gorgan\IEEEauthorrefmark{1}}
\IEEEauthorblockA{
\IEEEauthorrefmark{1}Computer Science Department, Technical University of Cluj-Napoca, Cluj-Napoca, Memorandumului 28, 
Romania\\
email: \{denisa.copandean, dorian.gorgan\}@cs.utcluj.ro\\
\IEEEauthorrefmark{3}Isaac Newton Group of Telescopes (ING), Apto. 321, 38700
Santa Cruz de la Palma, Canary Islands, Spain\\
Instituto de Astrofisica de Canarias (IAC), C/Via Lactea s/n,
E-38205 La Laguna, Spain\\
Departamento de Astrofisica, Universidad de La Laguna, E-38206 La Laguna, Tenerife, Spain\\
email: ovidiu.vaduvescu@gmail.com}}


\maketitle

\begin{abstract}
Near Earth Asteroids (NEAs) are discovered daily, mainly by few major surveys, nevertheless many of them remain unobserved for years, even decades. Even so, there is room for new discoveries, including those submitted by smaller projects and amateur astronomers. Besides the well-known surveys that have their own automated system of asteroid detection, there are only a few software solutions designed to help amateurs and mini-surveys in NEAs discovery. Some of these obtain their results based on the “blink” method in which a set of reduced images are shown one after another and the astronomer has to visually detect real moving objects in a series of images. This technique becomes harder with the increase in size of the CCD cameras. Aiming to replace manual detection we propose an automated pipeline prototype for asteroids detection, written in Python under Linux, which calls some 3rd party astrophysics libraries.
\end{abstract}


%
\IEEEpeerreviewmaketitle

\section{Introduction}
Near Earth objects (NEOs) are important topics of research and education in astrophysics and space sciences. Due to their small size, poor magnitude and very fast movement, NEOs discovery is a very difficult task, especially when the surveillance area is large. Besides their risk of collision with the Earth, these objects represent opportunities for mining resources, modeling new frontiers for emerging space industry. Monitoring the nearby space for near Earth asteroids (NEAs) is essential for the future of our planet. These studies have been geared first towards the main belt asteroids (MBAs), discovering more than 700,000 objects so far, as well as to the farthest asteroids of the Kuiper belt. Some research and surveys have been improved and directed towards NEAs smaller than 150 meters, the most difficult to detect and monitor. In the last couple of years, the NEAs discovery rate has been intensified due to the improved Catalina survey and to the new Pan-STARRS survey supported by the US Congress. While the major American surveys have increased discovery rate, in Europe there were just a few local initiatives led by some scientists without specific funding and dedicated facilities.

\begin{figure}[t!]
\centering
\includegraphics[width=3.4in]{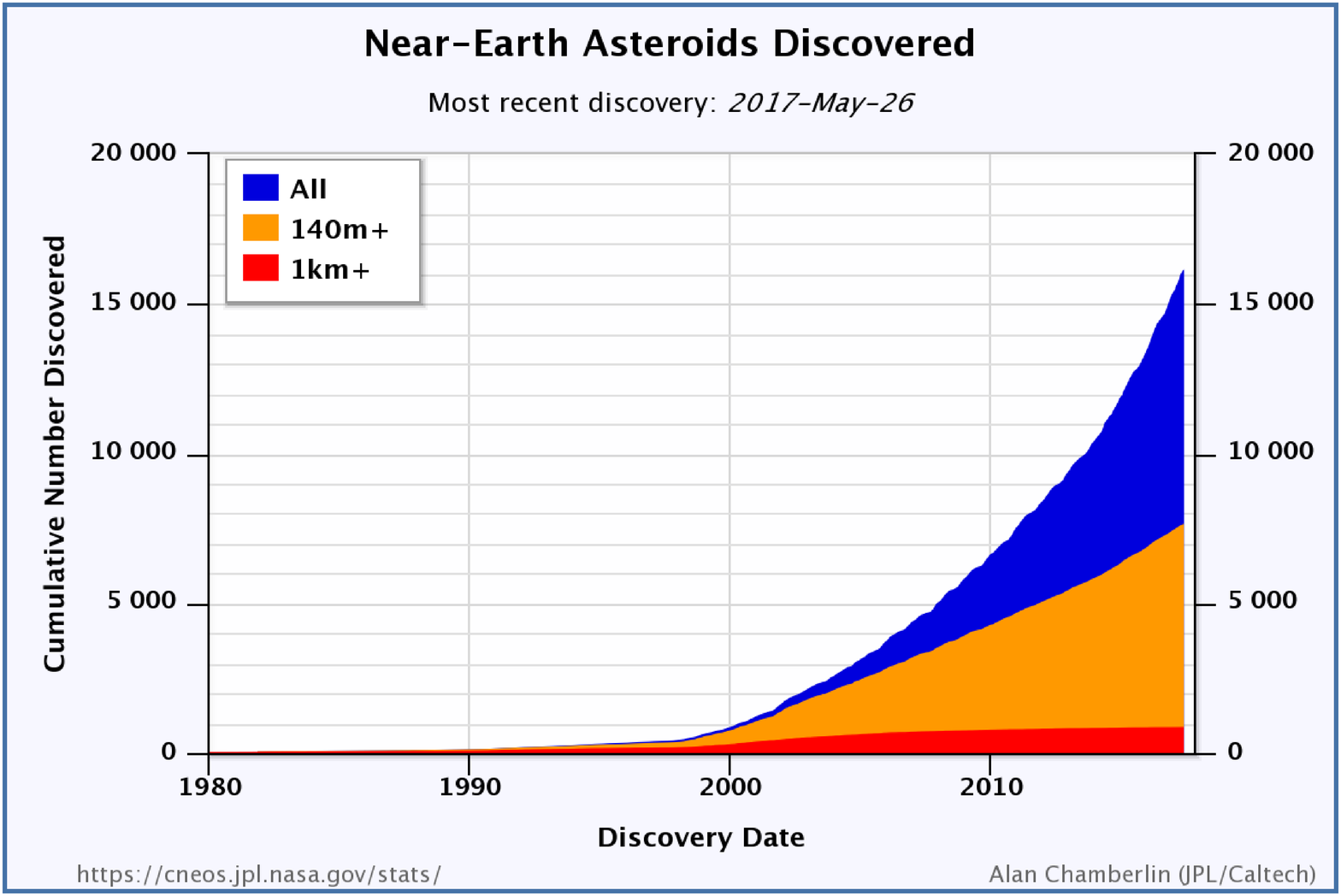}
\caption{NEA discoveries in time (https://cneos.jpl.nasa.gov/stats/totals.html)}
\label{fig:fig1}
\end{figure}

The European Near Earth Asteroids Research (EURONEAR) project has been contributing to this research since 2006 \cite{Vaduvescu2015}. EURONEAR includes only a small number of professional astronomers, involving more amateurs and students who promptly reduce the data, report discoveries and perform NEA recoveries. For the image reduction, THELI software is used (under Linux). This is a tool for the automated reduction of astronomical images \cite{Schirmer2013}, \cite{Erben2005} which can be used via a graphical user interface or by a set of command line based scripts. It subtracts the bias, twilight flat and corrects the known field distortion of the telescope’s prime focus field. Afterwards, the Astrometrica software, under Windows, developed by Herbert Raab is used in order to detect all moving sources via human blink \cite{Raab}. For the THELI reduced images to be used in Astrometrica, an intermediate manual step is performed to update the image headers to be supported by Astrometrica. Having this manual flow in mind, we propose an automated pipeline prototype for moving object detection. The prototype is a modular system written in Python language. Each module has a specific functionality and some of them are tightly coupled to some 3rd party astrophysics libraries.

The next section will go through some similar systems. Section 3 describes in more detail each of the modules and their dependences. The matching of identified objects is described by section 4. In section 5 we will present the major problems encountered and their solutions. The last section concludes on the achievements and experiments, presents the work currently in progress, and sketches some future research directions. 

\section{Related Works}
All of the major American surveys have written their own automated system for asteroids detection but virtually none of them is made public. The Near-Earth Asteroid Tracking (NEAT) program made the first fully automated system for controlling a remote telescope, acquiring wide-field images, and detecting NEOs \cite{Pravdo1999}. The WISE/NEOWISE Moving Object Pipeline Subsystem (WMOPS) was designed to detect Solar system objects in the WISE survey data as these data were processed in real-time, with particular emphasis on the detection of NEOs \cite{Cutri2011}.

Variable KD-Tree algorithms were used by the Pan-STARRS Moving Object Processing System \cite{Kubica2006}, \cite{Denneau2013}. These techniques were developed in collaboration with the incoming LSST (Large Synoptic Survey Telescope) \cite{Kub2007}.

Not only the major surveys developed automated system for moving object detection, but also some amateurs and small private surveys. A group of mainly amateurs search for NEAs in the TOTAS survey carried out with the ESA-OGS 1m telescope, lead by ESA \cite{Koschny2015}. Another group from Argentina developed such a system based on the profile of the each light source represented by FWHM (Full Width at Half Maximum) \cite{Allekotte2013}.

\section{Automated Moving Object Detection Prototype}
The current version of the proposed pipeline prototype is tightly coupled with the data obtained from the 2.5 meters diameter Isaac Newton Telescope (INT) located in La Palma, Canary Islands, Spain. At the prime focus of the INT, a wide field camera (WFC) is located, consisting of four CCDs 2k x 4k pixels, covering an L-shaped 34 arcmin x 34 arcmin field with a pixel scale of 0.33 arcsec pixel$^{-1}$.

\begin{figure}[t!]
\centering
\includegraphics[width=3.4in]{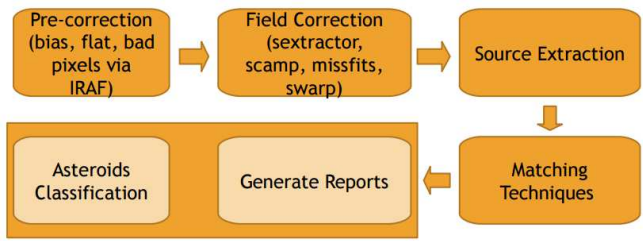}
\caption{Prototype Architecture}
\label{fig:fig2}
\end{figure}

Figure \ref{fig:fig2} presents the architecture of the prototype. The first two modules are based on a couple of external (3rd party) astrophysics libraries for image reduction and field correction. The last two modules automatically generate and submit reports of the new discovered objects under a specific format that is used by Minor Planet Center will be soon implemented.

\subsection{Pre-correction Module}
Any image taken by a CCD camera through a telescope will not give accurate information about the photometric distribution over a portion of the sky. Optical imperfections and the discrete nature of light itself lead to errors in the measured data. The first module corrects artifacts (bias, flat field) and instrumental defects (bad pixels) from the raw science images. It is embedded in Python and based on IRAF (Image Reduction and Analysis Facility) calls, which is a software system written and supported by the National Optical Astronomy Observatory (NOAO) in Tucson, Arizona, for the reduction and analysis of astronomical data. IRAF is distributed by NOAO, which is operated by the Association of Universities for Research in Astronomy (AURA) under a cooperative agreement with the National Science Foundation.

\begin{figure}[t!]
\centering
\includegraphics[width=3.0in]{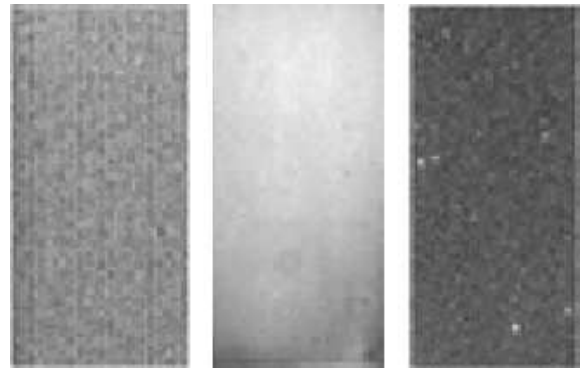}
\caption{Bias; Flat; Reduced Image}
\label{fig:fig3}
\end{figure}

\subsection{Field Correction Module}

\begin{figure*}[htbp]
\centering
\includegraphics[width=5.8in]{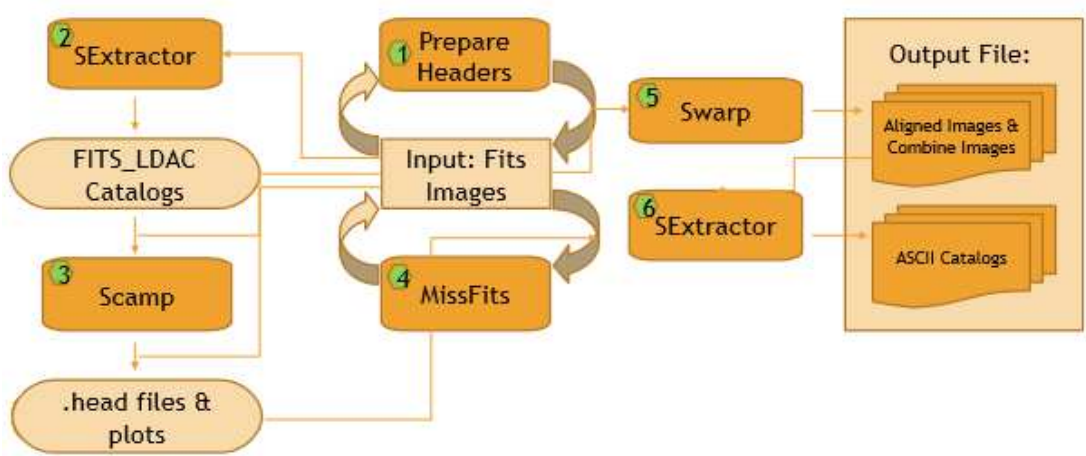}
\caption{Field Correction Module’s Scheme}
\label{fig:fig4}
\end{figure*}

Especially at the prime focus of any larger field telescope, substantial field distortions caused by the optical system and projection of a hemispherical field into a 2D image occur in the imaging process, which become difficult to perfectly align images for further processing and analysis. Field distortions increase with distance from the center of the image. This prototype solves the field distortion issue of the reduced images in a sequence of steps (Figure \ref{fig:fig4}).

This is the second module of the prototype, and it depends on few external astrophysics libraries. The first step, prepare headers, is needed to update image headers with some specific telescope information in order to be able to call the SExtractor in the second step. SExtractor is a software that builds a catalogue of objects from an astronomical image \cite{Bertin1996}. The output catalogs, in the FITS LDAC format, are used as input for the third step. In this step the SCAMP tool is used to obtain the shifting function for each pixel in order to correct field distortion \cite{Bertin2006}. Scamp reads SExtractor catalogs and computes astrometric and photometric solutions for any arbitrary sequence of FITS images in a completely automatic way. Its process is based on querying online catalogs of stars to find star references in the reduced images. The Scamp solution is stored in .head files and can be checked visually using plots files. All the new information obtained from Scamp is updated for each image using the MissFits software tool. Afterwards the SWarp tool is called in order to resample the images based on the shifting formula, thus correcting the field distortion in the image. Besides that, SWarp aligns and co-adds the reduced images of the same field of view. The combined images are needed in the next module for background subtraction. The last step of this module calls SExtractor again, but this time using the combined and resampled images, resulting in ASCII catalogs needed in the next module as well.

\subsection{Source Extraction Module}
Background subtraction of celestial sources is an important phase for detecting moving objects. Stars and galaxies are considered fixed sources, fully populating the background field. If they can be eliminated, what remains are the moving objects (asteroids, comets and other Solar System objects) and the noise. Two directions were approached to achieve background subtractions:
\begin{itemize}
\item \textit{Pixel based } - using star/galaxies subtraction. Subtraction operation at pixel level between resampled image and the specific combined image using IRAF – not good enough (due to variable sky and telescope tracking quality), see the Figure \ref{fig:fig5};

\begin{figure}[htbp]
\centering
\includegraphics[width=2.5in]{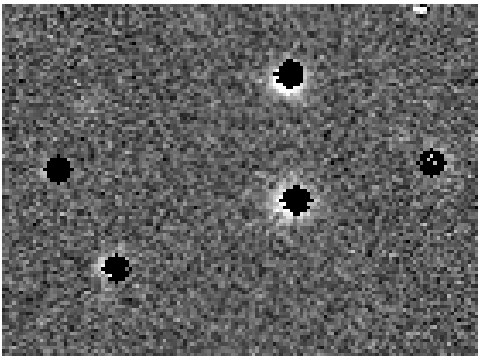}
\caption{Pixel Based Subtraction using IRAF}
\label{fig:fig5}
\end{figure}

\item \textit{Space Objects Based} - using sources in catalogs;
\begin{itemize}
\item represents the solution currently used in the prototype
\item moves the process entirely to Python which loads SExtractor ASCII catalog sources as Python objects
\item removes all fixed objects from individual catalogs that match sources from the combined catalog
\item the remaining objects are asteroids or noise and should be paired in the individual images
\end{itemize}
\end{itemize}

\section{Matching Module}
The matching module is based on the classic blink algorithm which is automatically processed in Python. When the space objects and catalogs are created in Python a couple of characteristics are stored as well, including full width at half maximum (FWHM). This parameter associated with a catalog can be interpreted as the clarity of the sky for the image that the catalog represents. The catalog of the image that have the best FWHM will be considered the pivot in the analyzed sequence of images (5 images). After that, for each source in the pivot, the algorithm will make pairs with sources from the others catalogs (pair – coupling a source from the pivot with a source from another catalog). In order for two sources to be paired, they need to be at a certain distance in known time interval, typical for most NEAs:
\begin{equation}
d = arccos[\sin\delta_{1} * \sin\delta_{2} + \cos\delta_{1} * \cos\delta_{2} * \cos(\alpha_{1} - \alpha_{2})]
\end{equation}

where:

$d -$ distance between two objects

$\delta_{i} -$ object declination

$\alpha_{i} -$ object right ascension\\

The actually matching is done in the second part of the module. If at least two pairs having the same source from the pivot comply with both the direction (based on the position angle $\omega$) and proper motion (represented by $\mu$), then we can assume that the object in cause is an asteroid.

\begin{figure}[t!]
\centering
\includegraphics[width=3.1in]{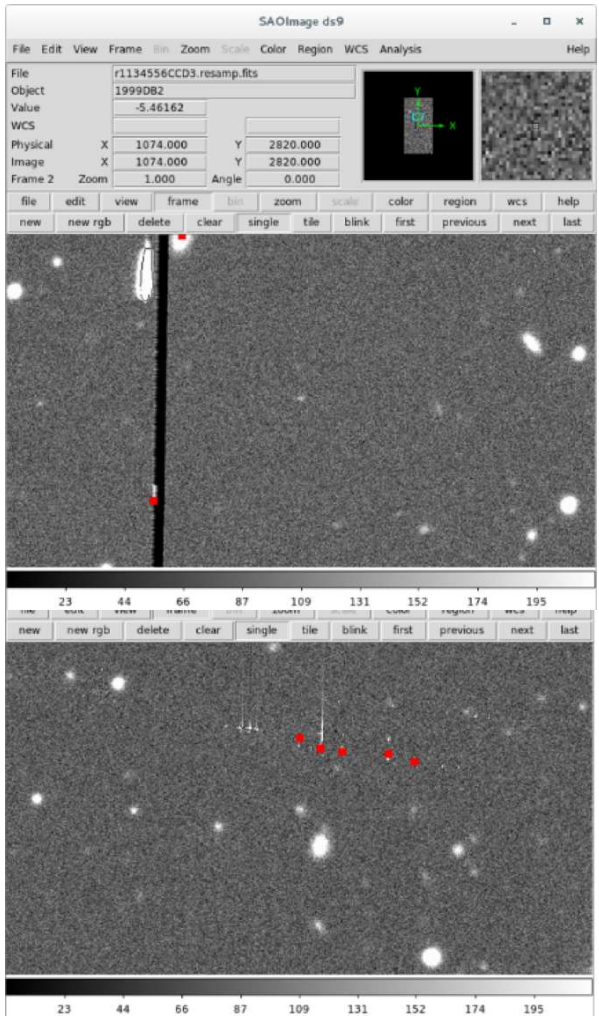}
\caption{Problems caused by bad pixels}
\label{fig:fig6}
\end{figure}

\begin{equation}
\mu = \frac{d}{\Delta UT}
\end{equation}

$\Delta UT -$ time frame between image acquisitions of the same observed field

\begin{equation}
\theta = arctan(\Delta \delta, \Delta \alpha)
\end{equation}

\begin{equation}
\omega = \theta + 2 * \pi, \theta < 0
\end{equation}

\begin{equation}
\omega = \theta, \theta \geq 0
\end{equation}
 
\begin{equation}
\mid \omega _{pair_{1}} - \omega _{pair_{2}} \mid < 10^{o} \ and \mid \mu _{pair_{1}} - \mu _{pair_{2}} \mid < 1"/min
\end{equation}

\section{Encountered Problems}
There were several types of problems encountered during prototype development. Among these are the following: 
\begin{itemize}
\item Caused by bad pixels - CCD detectors degrade over time, resulting in the occurrence of new "defective pixels", see Figure \ref{fig:fig6}
\begin{itemize}
\item Solution: generate new maps for "bad pixels" (the last dating from 2013)
\end{itemize}

\item Problems caused by interpolation – side effect of bad pixels correction, see Figure \ref{fig:fig7}. There are few bad columns resulted in imperfect correction, causing one bad pixel

\begin{figure}[t!]
\centering
\includegraphics[width=3.1in]{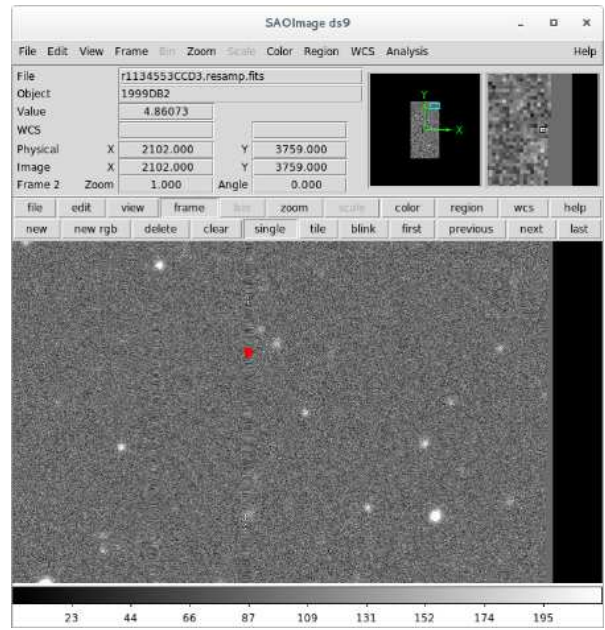}
\caption{Problems caused by interpolation}
\label{fig:fig7}
\end{figure}

\begin{figure}[b!]
\centering
\includegraphics[width=3.1in]{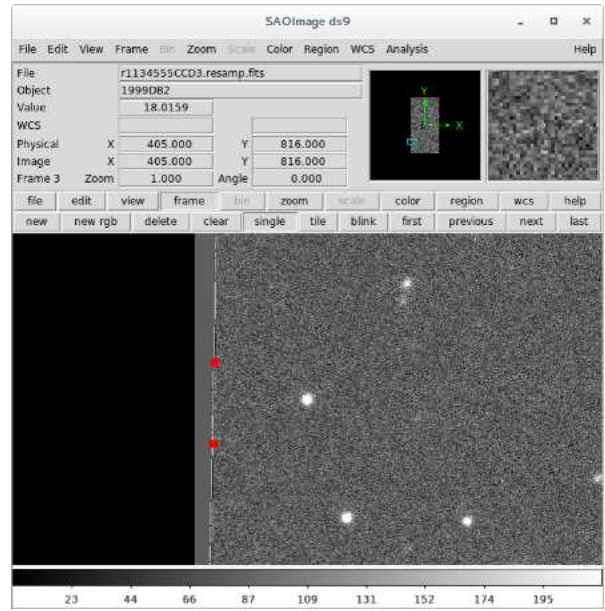}
\caption{Problems caused by image alignment}
\label{fig:fig8}
\end{figure}

\begin{itemize}
\item Solution: not yet implemented – visual analytics technique that will let the user to decide if it is a valid asteroid by showing them the use case in a graphical mode
\end{itemize}

\item Problems caused by image alignment - side effect of field distortion correction, see Figure \ref{fig:fig8}. The imperfect alignment of the camera causes few bad pixels in the margin of the rotated field

\begin{itemize}
\item Solution: SExtractor is an extremely powerful tool which, when properly configured, can bring out precious information about the objects identified in astronomical images. By activating certain parameters in the SExtractor configuration file, the resulting catalogs will contain a set of flags that characterize the object. Those flags are used in this solution. If they are different from zero, then surely the object is actually noise
\end{itemize}

\begin{figure}[htbp]
\centering
\includegraphics[width=3.1in]{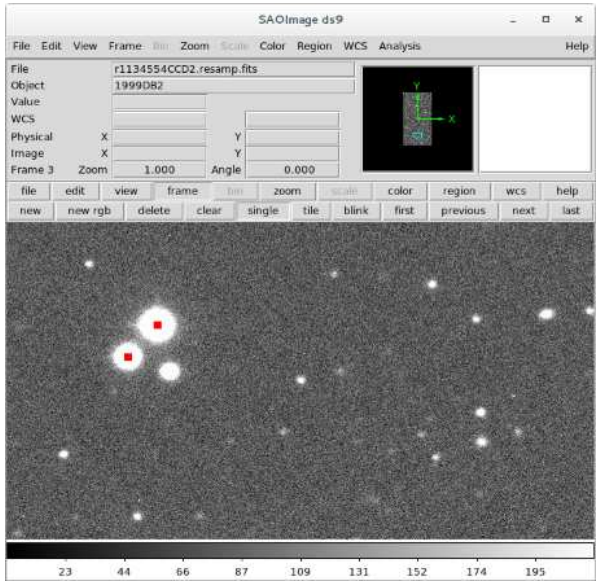}
\caption{Problems caused by star saturation}
\label{fig:fig9}
\end{figure}

\item Problems caused by star saturation, see Figure \ref{fig:fig9}. The very bright or saturated stars cause multiple detections close to the centre which can not be
properly subtracted
\begin{itemize}
\item Solution: based on SExtractor’s magnitude parameter, the very bright sources are eliminated
\end{itemize}

\begin{figure}[h!]
\centering
\includegraphics[width=3.1in]{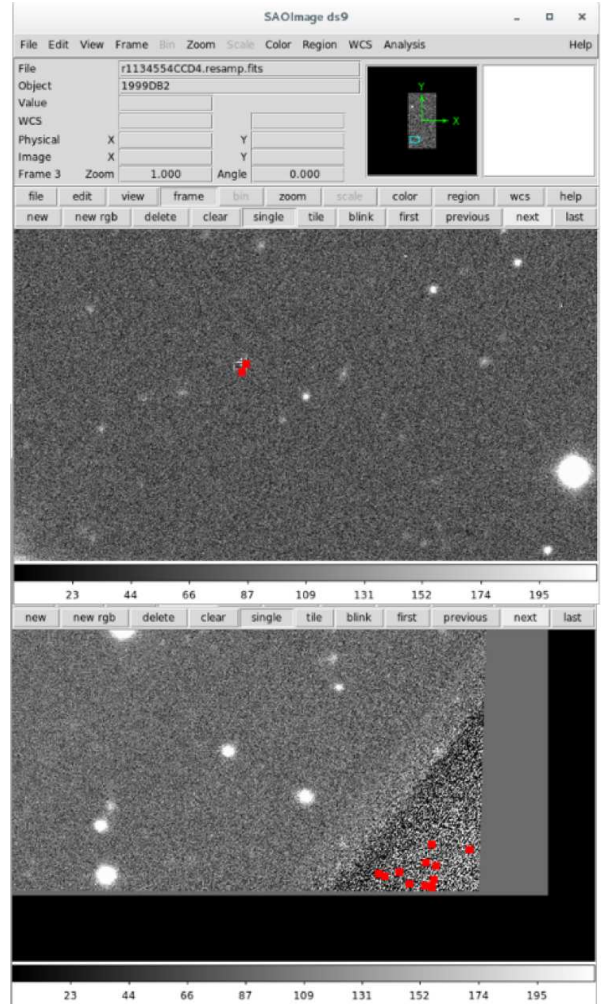}
\caption{Problems caused by noise}
\label{fig:fig10}
\end{figure}

\item Problems caused by noise – false detections or side effect of vignetting, see Figure \ref{fig:fig10}
\begin{itemize}
\item Solution: use a mask or the same as the solution proposed for the interpolation problem - a visual analytics technique will let the user to decide if it is a valid asteroid by showing them the use case in a graphical mode; sometimes, like in the case of the right picture in Figure \ref{fig:fig10}, it is possible to use a mask that will exclude the problematic area from the process
\end{itemize}

\item Problems caused by the short time frame between image acquisitions of the same observed field. If the image cadence is too fast, some of the moving objects will barely move and appear on the combined image and they will be interpreted as static sources or background, see Figure \ref{fig:fig11}

\begin{figure}[b!]
\centering
\includegraphics[width=2.5in]{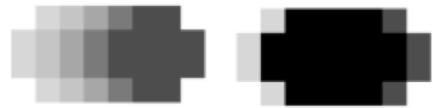}
\caption{Left: moving object in the sequence of images; Right: moving object in combined image}
\label{fig:fig11}
\end{figure}

\begin{itemize}
\item Solution: increase the time frame between image acquisitions of the same observed field by alternating pointing between four or more nearby survey fields
\end{itemize}

\end{itemize}

\section{Conclusions and Future Directions}
Until now, this prototype was only tested against a set of archives provided by the collaborative project EURONEAR. In this collection, the time frame between image acquisitions of the same observed field is very small and some of the asteroids identified by the human eye were flagged as part of the background in the prototype. Recently, a new set of dedicated data was provided and another round of testing began.
Medium and large sized telescopes (2-4 meters) are required to detect faint NEAs using the classic "blink" algorithm and in the near future it is planned to use this pipeline for such data. Even so, smaller telescopes (1 meter) can be used in the same purpose if the used process is based on various forms of "digital tracking" methods that require extensive computing resources. This is another direction that the prototype will approach in the future.
Having a modular structure, the prototype can be easy improved from the performance point of view. It will be migrated to cloud for  high performance computing. Taking advantage of parallel processing, the time will decrease considerably. Using Docker in the future cloud solution will bring a great value to any user that wants to use this prototype because the problems encountered in environment setup are eliminated. Some of the modules can be rewritten with image processing tools such as OpenCV. After that it can be further improved by the GPU using hardware accelerators tools such as CUDA or OpenCL (provided by NVIDIA).

\section*{Acknowledgment}
The first idea on such a work was proposed at the camp organized by the Romanian Astronomical Society of Meteors (SARM). The research and the experiments would not have been possible without getting involved in the activities of the EURONEAR group, and then through the ERASMUS mobility hosted by the Isaac Newton Group in La Palma, Canary Islands, Spain.



\bibliographystyle{IEEEtran}
\bibliography{IEEEabrv,paper}
%


\end{document}